\documentclass[journal=jctcce,manuscript=article,layout=twocolumn]{achemso}
\setkeys{acs}{etalmode=truncate,maxauthors=4}
\setkeys{acs}{articletitle=true}

\usepackage{chemformula} % Formula subscripts using \ch{}
\usepackage[T1]{fontenc} % Use modern font encodings
\usepackage{amsmath,amsfonts}
\usepackage{physics}
\usepackage{comment}
\usepackage{booktabs}
\usepackage{multirow}
\usepackage{textcomp}  % \textdegree symbol
\mciteErrorOnUnknownfalse
\usepackage{caption}
\usepackage{subcaption}
\usepackage{widetext}
\usepackage[version=4]{mhchem}
\usepackage{qcircuit}
\mathchardef\mhyphen="2D

\author{Joel Bierman}
\affiliation[Duke]{Department of Physics, Duke University}
\author{Yingzhou Li}
\affiliation[Fudan]{School of Mathematical Sciences, Fudan University}
\email{yingzhouli@fudan.edu.cn}
\author{Jianfeng Lu}
\email{jianfeng@math.duke.edu}
\affiliation[Duke]{Department of Mathematics, Duke University}
\alsoaffiliation{Department of Physics, Duke University}
\alsoaffiliation{Department of Chemistry, Duke University}

\title[Orbital Optimization]{Improving the Accuracy of Variational Quantum
Eigensolvers With Fewer Qubits Using Orbital Optimization}

\abbreviations{}

\begin{document}

% \begin{tocentry}
% 
%     Some journals require a graphical entry for the Table of Contents.
%     This should be laid out ``print ready'' so that the sizing of the
%     text is correct.
% 
%     Inside the \texttt{tocentry} environment, the font used is Helvetica
%     8\,pt, as required by \emph{Journal of the American Chemical
%     Society}.
% 
%     The surrounding frame is 9\,cm by 3.5\,cm, which is the maximum
%     permitted for  \emph{Journal of the American Chemical Society}
%     graphical table of content entries. The box will not resize if the
%     content is too big: instead it will overflow the edge of the box.
% 
%     This box and the associated title will always be printed on a
%     separate page at the end of the document.
% 
% \end{tocentry}

\begin{abstract}

Near-term quantum computers will be limited in the number of qubits on
which they can process information as well as the depth of the circuits
that they can coherently carry out. To-date, experimental demonstrations
of algorithms such as the Variational Quantum Eigensolver (VQE) have been
limited to small molecules using minimal basis sets for this reason. In
this work we propose incorporating an orbital optimization scheme into
quantum eigensolvers wherein a parameterized partial unitary
transformation is applied to the basis functions set in order to reduce
the number of qubits required for a given problem. The optimal
transformation is found by minimizing the ground state energy with respect
to this partial unitary matrix. Through numerical simulations of small
molecules up to 16 spin orbitals, we demonstrate that this method has the
ability to greatly extend the capabilities of near-term quantum computers
with regard to the electronic structure problem. We find that VQE paired
with orbital optimization consistently achieves lower ground state
energies than traditional VQE when using the same number of qubits and even frequently achieves lower ground state
energies than VQE methods using more qubits.

\end{abstract}

\section{Introduction}\label{sec:intro}

One of the main areas of research being conducted in quantum computing
today is exploring the extent to which near-term quantum computers can
be useful for solving practical problems. Any algorithm developed for
this purpose must fulfill three primary criteria: \textbf{1.} use as few
qubits as possible, \textbf{2.} minimize circuit depth, and \textbf{3.}
be robust to noise. One of the most promising problems for demonstrating
quantum advantage on near term quantum hardware is the electronic structure
problem~\cite{helgaker2014molecular}. The canonical approach to this
problem in quantum computing has been to use the second quantization
formulation, wherein we take the spatial coordinate representation of the
electronic structure Hamiltonian and project it onto a finite set of basis
functions. The choice of which basis to use ultimately determines how closely
the obtained energy levels using this truncated Hamiltonian will match those
of laboratory experimental results. Experimental results for demonstrating
quantum algorithms have so far been limited to representing small molecules
with minimal basis sets~\cite{McCaskey2019, PhysRevResearch.2.043140,
Kandala2017}. Such basis sets are useful for proof-of-concept demonstrations
and for benchmarking progress, but they do not represent results that would
match laboratory results well enough to be useful to a chemist. The ability
to move beyond these minimal basis sets will be an important step towards
demonstrating quantum advantage in computational chemistry. Doing so,
however, presents an obvious obstacle: Using larger basis sets increases
the qubit requirements for the simulation. Furthermore, many near-term
quantum algorithms developed for the electronic structure problem involve
the use of ansatz circuits with depth scaling polynomially with the size
of the spin orbital basis set.  Thus, increasing the size of the basis set
results in increased circuit depth as well.

Several methods have been proposed in recent years to make the
representation of the electronic structure Hamiltonian as compact and
resource-efficient on quantum computers as possible. These methods can be
roughly grouped into three categories: \textbf{1.} Classical
pre-processing of compact effective Hamiltonians, \textbf{2.} Orbital
optimizations interleaved between successive quantum eigensolver problems,
and \textbf{3.} post-processing to partially correct the basis set error.
Downfolded effective Hamiltonian techniques~\cite{DUCC_Hamiltonian,
unitarily_downfolded_hamiltonian, Claudino_2021} use a unitary
coupled-cluster ansatz operator to rotate the Hamiltonian in the full
orbital space, where the coupled-cluster amplitudes are solved for
classically. The transformed Hamiltonian is approximated according to a
second-order Baker-Campbell-Hausdorff expansion and projected onto a
chosen active space. Transcorrelated and explicitly correlated Hamiltonian
methods~\cite{CT-F12_hamiltonian_vqe, Transcorrelated_VarQITE} are
conceptually similar to downfolded methods, with the main difference being
that the similarity transformed applied to the Hamiltonian has an explicit
dependence on the coordinate space positions of the electrons. The purpose
of this is to efficiently capture the anti-correlation effects arising
from the Coulomb repulsion between electrons that would traditionally
require large basis set expansions. Orbital optimization methods share
some similarities to effective Hamiltonian methods in that they also apply
a similarity transformation to the Hamiltonian, but differ in how the
transformation parameters are found. Whereas effective Hamiltonian methods
solve for the transformation parameters in a pre-processing step, orbital
optimization methods\cite{OO-UCCD, SA-OO-VQE, PhysRevResearch.3.033230}
apply a parameterized unitary transformation to the Hamiltonian,
projecting the resulting parameterized Hamiltonian onto a chosen active
space and minimizing an objective function. The use of post-processing to
partially correct the error arising from the truncated basis set has also
been proposed. Virtual Quantum Subspace Expansion (VQSE)\cite{VQSE} is a method where
the ground state problem is first solved within a chosen active space
using an algorithm such as VQE. An improved estimate for the ground
state is then obtained by classically solving a generalized eigenvalue
problem over a contracted subspace spanned by single and double fermionic
excitation operators acting on the solution to the previous active space
problem. These excitation operators are allowed to include excitations to
the virtual space and thus contribute to a correction to the energy from
the limited active space solution. 

In this work we generalize the
OptOrbFCI~\cite{doi:10.1021/acs.jctc.0c00613} algorithm (developed in the
context of classical computing for settings in which classical
computational resources are limited) to the quantum computing setting in
which qubit counts and coherent circuit depth are limited resources.
OptOrbFCI is an orbital optimization method that applies a partial
unitary transformation to the set of basis functions, collapsing it to one
of a smaller size and introducing the elements of the matrix
representation of this transformation as additional parameters to be
optimized in the overall ground state search problem. An FCI solver is
used to find the ground state energy in a reduced basis. Extending
OptOrbFCI to the quantum computing setting corresponds to replacing the
FCI solver subroutine with one of several quantum eigensolvers such as the
Variational Quantum Eigensolver (VQE)~\cite{Peruzzo2014}, Quantum
Imaginary Time Evolution (QITE)~\cite{Motta2020, McArdle2019}, or Quantum
Monte Carlo~\cite{Huggins2022}. In this work, we pair the orbital
optimization subroutine with VQE, calling the resulting overall method
OptOrbVQE. We find that OptOrbVQE consistently achieves lower ground state
energy compared to standard VQE  methods when using the same number of
qubits. Higher accuracy results are also achieved while simultaneously
using fewer qubits than these methods in several instances.

The rest of the paper is organized as follows. In \S\ref{sec: VQE} we give
a brief overview of the main method for computing ground states in quantum
computing, VQE. In \S\ref{sec: OptOrb} we propose the orbital optimization
approach in the setting of variational quantum eigensolvers to reduce
the resource requirement of qubits. In \S\ref{sec: Numerical Results} we
benchmark OptOrbVQE on several small molecules. In \S\ref{sec: Discussion
and Conclusions} we discuss the results and potential directions of future
research.

\section{Variational Quantum Eigensolver}\label{sec: VQE}

One promising method for computing the ground state of chemical systems on
near-term quantum hardware is the Variational Quantum Eigensolver (VQE).
The method begins by formulating the electronic structure Hamiltonian in
the second quantization as
\begin{equation}\label{eq: fermionic hamiltonian}
    \hat{H} = \sum _{p,q = 1}^{M} h_{pq}\hat{a}^{\dagger}_{p}\hat{a}_q
    + \frac{1}{2}\sum_{p,q,r,s = 1}^{M} v_{pqrs}
    \hat{a}^{\dagger}_{p}\hat{a}^{\dagger}_{q}\hat{a}_{s}\hat{a}_{r},
\end{equation}
where $h_{pq}$ and $v_{pqrs}$ are the one and two-electron integrals as
in Eq.~\eqref{eq: one body integral} and Eq.~\eqref{eq: two body integral}
over our set of $M$ basis functions $\{\psi_1, \psi_2, ..., \psi_M\}$.
\begin{equation}\label{eq: one body integral}
    h_{pq} = \int d\textbf{x}_{1}
    \psi^*_{p}(\textbf{x}_{1})h(\textbf{x}_{1})\psi_{q}(\textbf{x}_{1})
\end{equation}
\begin{equation}\label{eq: two body integral}
\begin{split}
    v_{pqrs} = \int d\textbf{x}_{1}&d\textbf{x}_{2}\psi_{p}^*(\textbf{x}_1)\psi^*_{q}(\textbf{x}_2)\\
    & \times v(\textbf{x}_{1}, \textbf{x}_{2})\psi_{s}(\textbf{x}_{2})\psi_{r}(\textbf{x}_{1})
\end{split}
\end{equation}
This fermionic Hamiltonian can be mapped to a qubit Hamiltonian of the form in
Eq.~\eqref{eq: qubit Hamiltonian} by using one of several known mapping schemes
such as Jordan-Wigner, Parity, or Bravyi-Kitaev~\cite{RevModPhys.92.015003}.
\begin{equation}\label{eq: qubit Hamiltonian}
    \hat{H} = \sum_{i} h_{i}\hat{P}_{i}
\end{equation}
Here $\hat{P}_{i}$ are tensor products of local Pauli operators acting on
a register of qubits. The quantum computer can measure expectation values
of these Pauli operators and a classical computer computes their weighted
sum. The wavefunction is parameterized as $\ket{\psi(\pmb{\theta})} =
\hat{U}(\pmb{\theta})\ket{\psi_{\text{ref}}}$, where $\ket{\psi_{\text{ref}}}$
is an initial reference state of our choice and $\hat{U}(\pmb{\theta})$
is a parameterized quantum ansatz circuit. Using the variational principle,
the ground state search problem can be formulated as the minimization problem:
\begin{equation}
    \min_{\pmb{\theta}}\bra{\psi_{\text{ref}}}\hat{U}^{\dagger}(\pmb{\theta})\hat{H}\hat{U}(\pmb{\theta})\ket{\psi_{\text{ref}}}
\end{equation}
A quantum computer prepares the wavefunction and measures the Hamiltonian
expectation value, then passes this value to a classical gradient-free
optimization subroutine, which returns a new value for the parameters.
This process repeats until the stopping condition of the optimizer is reached.

\section{Optimal Orbital VQE}\label{sec: OptOrb}

Let us now introduce the orbital optimization in the VQE setting, motivated
by a similar scheme in the classical setting as the OptOrbFCI algorithm
proposed by two of the authors.~\cite{doi:10.1021/acs.jctc.0c00613} If our
set of basis functions has size $M$, then this will require the use of $M$
qubits if no techniques to reduce this count are employed. Suppose we have
access to a quantum computer with only $N < M$ qubits or that we are using
an ansatz circuit that scales with the number of qubits in such a way that
we are limited to calculations using $N$ qubits. We thus have to restrict
to a Hamiltonian with only $N$ spin orbitals by applying a partial unitary
transformation for the basis change, which we represent using a $M \times N$
real partial unitary matrix $\hat{V}$. The basis functions will transform
according to
\begin{equation}\label{eq: basis func transform}
    \Tilde{\psi_{i}} = \sum_{j}^{M} \hat{V}_{ji}\psi_{j}
\end{equation}
This corresponds to the one and two body integrals transforming according
to Eq.~\eqref{one body integral transform} and Eq.~\eqref{two body integral
transform}.

\begin{equation}\label{one body integral transform}
    \Tilde{h}_{p^{\prime}q^{\prime}} = \sum_{p,q = 1}^{M} h_{pq}\hat{V}_{pp^{\prime}}\hat{V}_{qq^{\prime}}
\end{equation}
\begin{equation}\label{two body integral transform}
    \Tilde{v}_{p^{\prime}q^{\prime}r^{\prime}s^{\prime}} = \sum_{p,q,r,s = 1}^{M} v_{pqrs} \hat{V}_{pp^{\prime}}\hat{V}_{qq^{\prime}}\hat{V}_{ss^{\prime}}\hat{V}_{rr^{\prime}}
\end{equation}
The ground state energy is now a function of not only the ansatz parameters
$\pmb{\theta}$, but the partial unitary matrix $\hat{V}$ as well. The ground
state search problem is now a minimization problem over both the space of
ansatz parameters and the space of all real partial unitary matrices of
dimension $M \times N$:

\begin{equation}\label{eq: total minimization problem}
    \min_{\substack{\pmb{\theta}\\
    \hat{V}\in\mathcal{U}(M,N)}} \bra{\psi_{\text{ref}}}\hat{U}^{\dagger}(\pmb{\theta})\Tilde{H}(\hat{V})\hat{U}(\pmb{\theta})\ket{\psi_{\text{ref}}}
\end{equation}
where
\begin{equation}
    \mathcal{U}(M,N) = \{\hat{V}\in\mathbb{R}^{M\times N} | \hat{V}^T\hat{V} = I_{N}\}
\end{equation}
The transformed Hamiltonian as a function of $\hat{V}$ is given by:
\begin{equation}
\begin{split}
    &\Tilde{H}(\hat{V}) = \sum_{p^{\prime}, q^{\prime} = 1}^{N} \sum_{p,q = 1}^{M} h_{pq}\hat{V}_{pp^{\prime}}\hat{V}_{qq^{\prime}}\hat{a}^{\dagger}_{p^{\prime}}\hat{a}_{q^{\prime}}\\
    &+ \frac{1}{2}\sum_{p^{\prime},q^{\prime},r^{\prime},s^{\prime}=1}^{N}\sum_{p,q,r,s=1}^{M}v_{pqrs}\hat{V}_{pp^{\prime}}\hat{V}_{qq^{\prime}}\hat{V}_{ss^{\prime}}\hat{V}_{rr^{\prime}}\hat{a}^{\dagger}_{p^{\prime}}\hat{a}^{\dagger}_{q^{\prime}}\hat{a}_{s^{\prime}}\hat{a}_{r^{\prime}}
\end{split}
\end{equation}
where the primed and unprimed indices index the transformed and original
basis wavefunctions, respectively. (\textit{e.g.} $\hat{a}^{\dagger}_{p}$
is the fermionic creation operator corresponding to spin-orbital
$\psi_{p}$ and $\hat{a}^{\dagger}_{p^{\prime}}$ is the fermionic
creation operator corresponding to the transformed spin-orbital
$\tilde{\psi}_{p^{\prime}}$.) This fermionic Hamiltonian can then be mapped
to a weighted sum of Pauli string operators acting on qubits. We leave
the Hamiltonian expressed in terms of fermionic operators to emphasize
that the method is independent of the mapping chosen. The expectation values
$\bra{\psi_{\text{ref}}}\hat{U}^{\dagger}(\pmb{\theta})\hat{a}^{\dagger}_{p^{\prime}}\hat{a}_{q^{\prime}}\hat{U}(\pmb{\theta})\ket{\psi_{\text{ref}}}$
and
$\bra{\psi_{\text{ref}}}\hat{U}^{\dagger}(\pmb{\theta})\hat{a}^{\dagger}_{p^{\prime}}\hat{a}^{\dagger}_{q^{\prime}}\hat{a}_{s^{\prime}}\hat{a}_{r^{\prime}}\hat{U}(\pmb{\theta})\ket{\psi_{\text{ref}}}$
are the 1-RDM and 2-RDM elements $^{1}D^{p^{\prime}}_{q^{\prime}}$ and
$^{2}D^{p^{\prime},q^{\prime}}_{r^{\prime},s^{\prime}}$, respectively.
These quantities are (after being mapped to qubit operators) measured on a
quantum computer with respect to the ansatz state $\ket{\psi(\pmb{\theta})}
= \hat{U}(\pmb{\theta})\ket{\psi_{\text{ref}}}$ in the same fashion as
conventional VQE.

It is important to note that the optimization problem in Eq.~\eqref{eq:
total minimization problem} consists of two distinctly different types of
parameters subject to different types of constraints: the partial unitary
$\hat{V}$ and the vector $\pmb{\theta}$ (which typically consists of real
numbers subject to some bounds). Thus, it is natural to treat the two sets
of variables separately.  In this work we adopt the procedure originally
proposed by OptOrbFCI in the classical setting. The minimization problem in
Eq.~\eqref{eq: total minimization problem} is divided into two subproblems:
minimizing the energy with respect to $\hat{V}$ (keeping $\pmb{\theta}$ fixed)
and minimizing the energy with respect to $\pmb{\theta}$ (keeping $\hat{V}$
fixed). We alternate between these two subproblems until some stopping
criterion is reached. Because this algorithm involves two minimization
subproblems (each with their own iteration number counter) that are both
repeated multiple times (where this number of times is associated with an
additional ``outer loop'' iteration number counter), we specify which indices
are used to denote which type of iteration counter throughout this paper in
order to reduce any ambiguity:
\begin{itemize} 
\item $l$ will be used to denote the iteration number within the
minimization with respect to $\hat{V}$;
\item $m$ will be used to denote the iteration number within the minimization
with respect to $\pmb{\theta}$ (the same as what is typically referred to as
the iteration number within the context of VQE without orbital optimization);
\item $n$ will be used to denote the outer loop iteration number
(\emph{i.e.} the number of times the minimization subproblem with respect
to $\hat{V}$ has been conducted so far). 
\end{itemize}
The superscript \emph{opt} will be used to denote the optimal point for
each of the minimization subproblems within a given outer loop iteration.
The OptOrbVQE algorithm can be summarized as follows:
\begin{enumerate}
    \item Set the outer loop iteration number $n = 0$ and choose an
    initial partial unitary transformation $\hat{V}_{n=0, l=0}$ and
    initial VQE parameters $\pmb{\theta}_{n=0,m=0}$. Choose an outer loop
    stopping tolerance $\epsilon_{outer}$.
    \item On a classical computer, calculate the transformed Hamiltonian
    $\Tilde{H}(\hat{V}_n)$ and use one of several known mappings to
    generate the corresponding transformed qubit Hamiltonian.
    \item Initialize the ansatz state as
    $\hat{U}(\pmb{\theta}_{n,m=0})\ket{\psi_{\text{ref}}}$ and perform VQE on a
    quantum computer to obtain $\pmb{\theta}^{\emph{opt}}_n$ and the
    estimated ground state energy $E(\hat{V}_{n=n,l=0},
    \pmb{\theta}^{\emph{opt}}_{n})$.
    \item If $\lvert E(\hat{V}_{n-1,l=0}, \pmb{\theta}_{n-1}^{\emph{opt}})
    - E(\hat{V}_{n,l=0}, \pmb{\theta}^{\emph{opt}}_n)\rvert <
    \epsilon_{outer}$, halt the algorithm and return
    $E(\hat{V}_{n=n,l=0})$,
    $\hat{U}(\pmb{\theta}^{\emph{opt}}_{n})\ket{\psi_{\text{ref}}}$, and
    $\hat{V}_{n,l=0}$ as the optimal quantities of interest. Else,
    continue to next step.
    \item On a quantum computer, measure the 1-RDM and 2-RDM elements with
    respect to the state
    $\hat{U}(\pmb{\theta}^{\emph{opt}}_n)\ket{\psi_{\text{ref}}}$.
    \item Initialize the partial unitary as $\hat{V}_{n,l=0}$ and perform
    the minimization subproblem in Eq.~\eqref{eq: total minimization problem}
    with respect to $\hat{V}$ (using the 1- and 2-RDM tensors from the
    previous step) to obtain $\hat{V}_{n}^{\emph{opt}}$.
    \item Set $\hat{V}_{n+1,l=0} = \hat{V}_{n}^{\emph{opt}}$ and
    $\pmb{\theta}_{n+1,m=0} = \pmb{\theta}^{\emph{opt}}_n$. Optionally, a
    small random perturbation can be added to $\hat{V}_{n+1,l=0}$ to avoid
    shallow local minima.
    \item Set $n = n + 1$ and repeat steps 2-8.
\end{enumerate}

There are a few clear initializations $\hat{V}_{n=0,l=0}$ and $\hat{V}_{n,l=0}$
that can be used in this algorithm. Throughout this work, we choose
$\hat{V}_{n=0,l=0}$ to be the permutation matrix that selects $N$ spin
orbitals from the starting basis with the lowest Hartree-Fock energy ordered
by ascending energy. This is equivalent to starting with a large basis, but
restricting the active space to these $N$ spin orbitals. This is not the only
initialization that could be used, but it is an intuitive one. In general,
we can take any $M\times N$ real matrix $A$ and project it onto one which
is a partial unitary through the orthonormalization function:
\begin{equation}
    \mathrm{orth}(A) = AQ\Lambda^{-\frac{1}{2}}Q^{\dagger}
\end{equation}
where $Q$ and $\Lambda$ together are a solution of the diagonalization equation
$A^{\dagger}A = Q\Lambda Q^{\dagger}$. We could, for instance,
orthonormalize a matrix whose elements are sampled from a random
distribution of our choice. The normal distribution or the uniform
distribution over some interval would be natural choices. If $\hat{P}$ is
the permutation matrix used in this work, then one alternative choice for
$\hat{V}_{n=0,l=0}$ would be $\mathrm{orth}(\hat{P} + \text{Rand}(M,N))$,
where $\text{Rand}(M,N)$ is a random $M \times N$ matrix. Throughout this
paper, the partial unitary $\hat{V}_{n+1,l=0}$ in step 7 of the algorithm
is chosen to be $\mathrm{orth}(\hat{V}_{n}^{opt} + \text{Rand}(M,N))$,
with the elements of $\text{Rand}(M,N)$ in this instance being sampled
from the normal distribution centered about mean 0 with a standard
deviation 0.01. The random perturbation matrix is added to help the method
avoid getting trapped in shallow local minima.

We end this section by noting the differences between this proposed
method and specific examples of methods in categories mentioned in the
introduction. \textbf{1.} In contrast to explicitly correlated and
downfolded Hamiltonian parameters, where the similarity transformation
parameters are found as a pre-processing step according to a pre-defined
set of equations or chemical intuition, OptOrbVQE (like other orbital
optimization methods) finds the optimal parameters by minimizing an
objective function. \textbf{2.} Many of the techniques referenced in the
introduction such as the DUCC~\cite{DUCC_Hamiltonian} and CT-F12
Hamiltonians~\cite{CT-F12_hamiltonian_vqe},
QDSRG~\cite{unitarily_downfolded_hamiltonian},
OO-UCC~\cite{OO-UCCD},SA-OO-VQE~\cite{SA-OO-VQE}, and quantum
CASSCF~\cite{PhysRevResearch.3.033230} use a similarity transformation
which takes the form of a chemically-motivated ansatz. The DUCC, CT-F12,
and QDSRG methods further approximate the transformed Hamiltonian
according to a second-order expansion. In OptOrbVQE, the similarity
transformation is not constrained by the form of an ansatz and can take
the form of a general partial unitary. This partial unitary matrix is then
applied directly to the one- and two-body integral tensors over the full
orbital space, removing the necessity of any approximations. In the other orbital
optimization techniques mentioned in the introduction, such as OO-UCC,
SA-OO-VQE, and quantum CASSCF, the use of a unitary transformation over
either the full orbital space or a subset of it necessitates the
partitioning of the full orbital space into core, active, and virtual
subspaces in order to reduce the problem to a manageable size. The orbital optimization subproblem in OptOrbVQE
is more flexible, with the choice of active space being determined automatically according to the minimization of an objective function. The removal of core or
virtual orbitals as a pre-processing step can be employed, but it is not
necessary.

\section{Numerical Results}\label{sec: Numerical Results}

Our implementation of the OptOrbVQE algorithm is a combination of
in-house code and code from the open source packages Qiskit~\cite{Qiskit}
(Qiskit Nature 0.3.2, Qiskit Aer 0.10.4, and Qiskit Terra 0.20.0) and
PyTorch~\cite{NEURIPS2019_9015} 1.11.0. The method of finding the optimal
$\hat{V}$ with fixed $\pmb{\theta}$ is the same as that used in the OptOrbFCI
proposal paper: a projection method with alternating Barzilai-Borwein
stepsize~\cite{doi:10.1137/16M1098759}. The code for this optimizer was
developed in-house using several tensor functionalities of PyTorch. We
choose to use PyTorch for several reasons: \textbf{1.} We find that it
has an efficient \emph{einsum} implementation which greatly speeds up the
computation of Eq.~\eqref{eq: total minimization problem}. \textbf{2.} It has
support for automatic differentiation, which enables efficient computation
of the gradient of Eq.~\eqref{eq: total minimization problem} with respect
to $\hat{V}$ in the projection method. \textbf{3.} It offers support for GPU
acceleration, which can speed up the calculation significantly, especially
for larger starting basis sets. The subproblem of minimizing the energy with
respect to $\pmb{\theta}$ uses Qiskit's VQE implementation.

\subsection{Minimal Qubit Usage}\label{sec: Minimal Qubit Usage}

In this section we investigate the ground state accuracy achievable by
OptOrbVQE when using the same number of spin orbitals as a minimal basis
set. We then compare the results to VQE and FCI simulations using basis sets of
the same size or larger. Ideally, we would only compare OptOrbVQE to VQE because
this is a more appropriate comparison than classical FCI methods. However, we
find that simulating VQE in Qiskit is much more computationally expensive than
carrying out an FCI problem of the same size using PySCF. Thus, FCI results
are a convenient stand-in for VQE results that would be computationally
infeasible. The assumption here is that the FCI ground state energy serves
as a lower bound for what is achievable by VQE. In the best-case scenario
where a sufficiently powerful ansatz is used and VQE achieves convergence
to the global minimum, these values would closely match.

The classical optimizer used in VQE subproblem instances in this section is
L-BFGS-B~\cite{doi:10.1137/0916069}. We use Qiskit's \emph{AerSimulator}
in combination Qiskit's \emph{AerPauliExpectation} algorithm to compute
expectation values of both the molecular Hamiltonian and the observables
involved in computing the 1 and 2-RDM. This combination yields ideal,
noiseless results. Thus, these simulations serve to test the ability of
the OptOrbVQE algorithm to converge under ideal conditions, but not its
robustness to noise. We defer a study of the robustness to noise of the
method to \S\ref{sec: Robustness to Noise}. The stopping tolerances for both
the orbital rotation subproblem and the OptOrbVQE algorithm as a whole are
set to $10^{-5}$. The maximum outer loop iteration number is set to 19 so
that the VQE subproblem is run at most 20 times.

\subsubsection{\ch{H4}}\label{sec: H4}

We begin by presenting classically-simulated results for \ch{H4}, a toy
model which consists of 4 hydrogen atoms arranged in a square with an H-H
distance of 1.23 \AA. The ansatz used is 2-UCCSD~\cite{Romero_2018}. In Qiskit,
one has the ability to repeat a base ansatz circuit $n$ times to produce a
more expressive ansatz. When we refer to $n$-UCCSD, we mean an ansatz which
consists of the base UCCSD ansatz repeated $n$ times in this fashion. Using
$n$-UCCSD has the effect of increasing both the circuit depth and the number
of independent parameters by a factor of $n$ over UCCSD. We find that two
repetitions are necessary for VQE in the STO-3G basis to converge to within
the chemical accuracy of the FCI value (calculated using PySCF~\cite{Sun2018}
2.0.1) in the same basis for the \ch{H4} toy model.

\begin{figure}[htb]
    \centering
    \includegraphics[width=\linewidth]{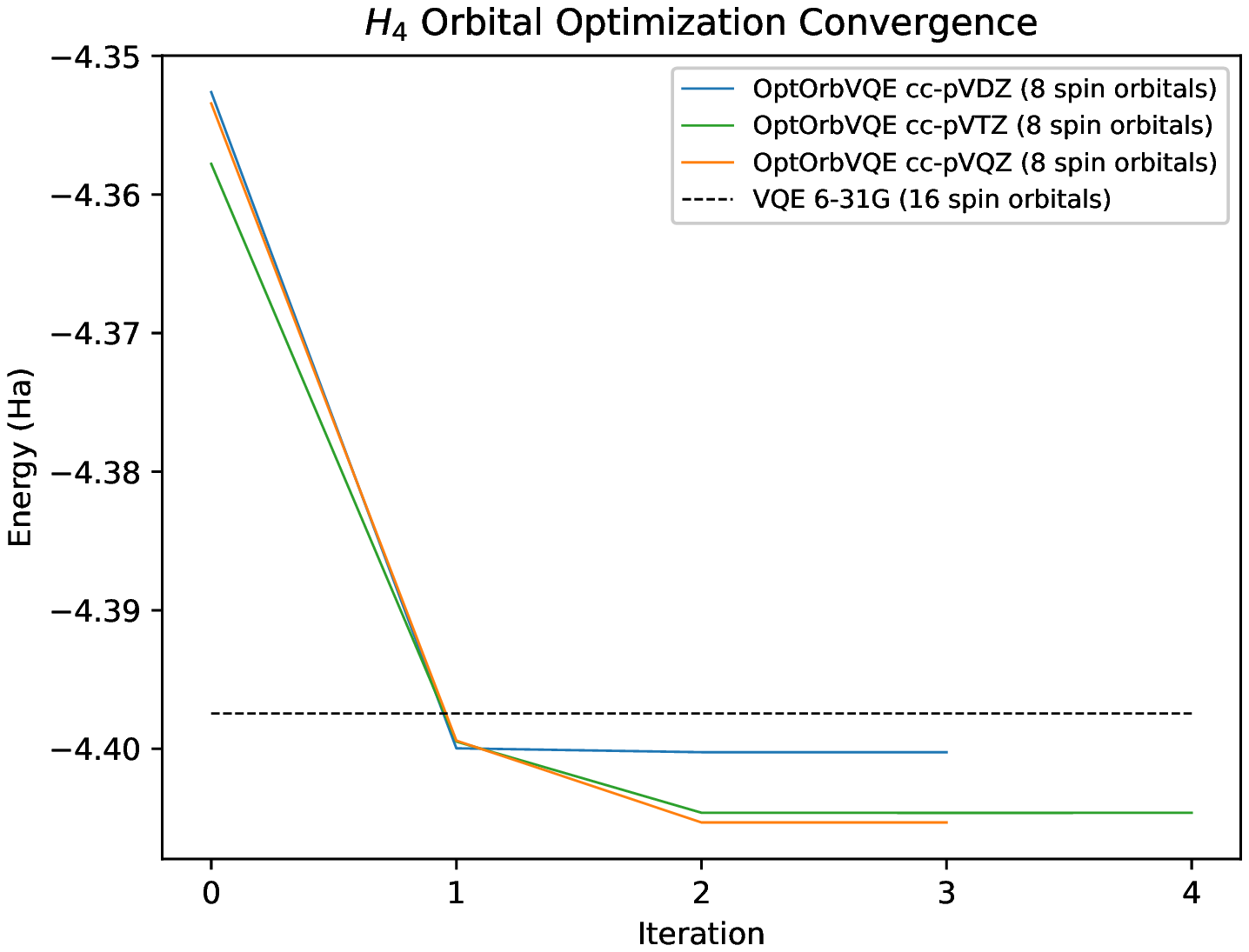}
    \caption{Convergence of OptOrbVQE as a function of the outer loop
    iteration number for \ch{H4} at the near-equilibrium \ch{H}-\ch{H} distance of
    1.23~\AA.}
    \label{fig: H4_equilibrium}
\end{figure}
We set the number of spin orbitals to be 8 for \ch{H4}, the number of spin
orbitals for this system in the minimal STO-3G basis set. Fig.~\ref{fig:
H4_equilibrium} illustrates the convergence of the OptOrbVQE algorithm as a
function of the outer loop iteration number for various starting basis sets. We
compare the results to that obtained from VQE in the 6-31G basis using 2-UCCSD
as the ansatz. Under these conditions, VQE is using 16 qubits. Despite the
fact that OptOrbVQE is using half the number of qubits as VQE, we find
that it achieves a lower ground state energy for all the starting basis
sets used. This lower energy is achieved after just the $n = 1$ outer loop
iteration, which corresponds to carrying out the orbital rotation subroutine
once and the VQE subroutine twice. The energy is lowered further when cc-pVTZ
and cc-pVQZ are used as starting basis sets with further iterations.

\subsubsection{\ch{LiH}}\label{sec: LiH}

For \ch{LiH}, we use 1-UCCSD as the ansatz. We set the number of spin orbitals
for OptOrbVQE to be 12, the number of spin orbitals for this system in
the minimal STO-3G basis set. We compute the ground state energy at the
near-equilibrium Li-H distance of 1.595 \AA~as well as the binding curve
of \ch{LiH}.

\begin{figure}[htb]
    \centering
    \includegraphics[width=\linewidth]{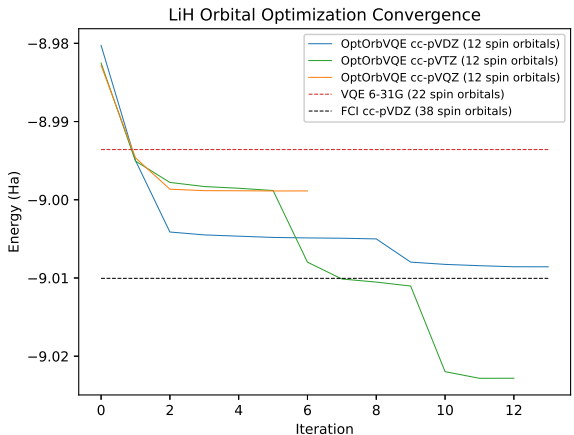}
    \caption{Convergence of OptOrbVQE as a function of the outer loop
    iteration number for \ch{LiH} at the near-equilibrium bond distance of
    1.595 \AA.}
    \label{fig:LiH_equilibrium}
\end{figure}

\begin{figure}[htb]
    \centering
    \includegraphics[width=\linewidth]{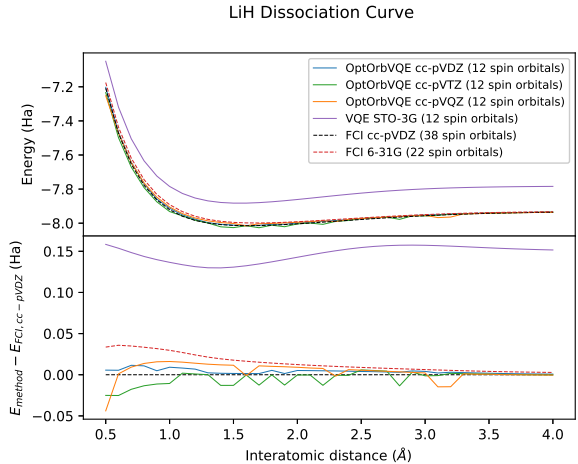}
    \caption{Top: Dissociation curve of \ch{LiH}. Bottom: Difference of
    energy relative to FCI (cc-pVDZ).}
    \label{fig:LiH_binding_curve}
\end{figure} 

Fig.~\ref{fig:LiH_equilibrium} illustrates the convergence of OptOrbVQE
as a function of the outer loop iteration number. We find that OptOrbVQE
achieves a lower energy than VQE in the 6-31G basis after the $n = 1$
iteration. The energy is further improved with additional iterations. In
particular, OptOrbVQE using cc-pVTZ as the starting basis surpasses the FCI
energy in the cc-pVDZ basis at the $n = 7$ iteration. OptOrbVQE starting
from the cc-pVDZ basis also approaches, but does not surpass this value. We
also note that starting from a larger basis does not always result in a
more accurate value, as can be seen from OptOrbVQE (cc-pVQZ starting basis)
not achieving the same accuracy as the other two starting basis sets.

Fig.~\ref{fig:LiH_binding_curve} illustrates the results obtained for the
binding curve of \ch{LiH}. We can see that OptOrbVQE easily outperforms VQE
using the same number of qubits. OptOrbVQE consistently achieves an energy
lower than the FCI energy in the 6-31G basis. OptOrbVQE also often achieves
an energy lower than the FCI energy in the cc-pVDZ basis, although this is
not guaranteed and sometimes fails to do so.

\subsubsection{\ch{BeH2}}\label{sec: BeH2}

In this section we test OptOrbVQE on \ch{BeH2}, a linear molecule with a
near-equilibrium Be-H bond distance of 1.3264 \AA. 1-UCCSD is the ansatz
used. The number of spin orbitals used by OptOrbVQE is set to 14, the number
of spin orbitals for this system in the minimal STO-3G basis.

\begin{figure}[htb]
    \centering
    \includegraphics[width=\linewidth]{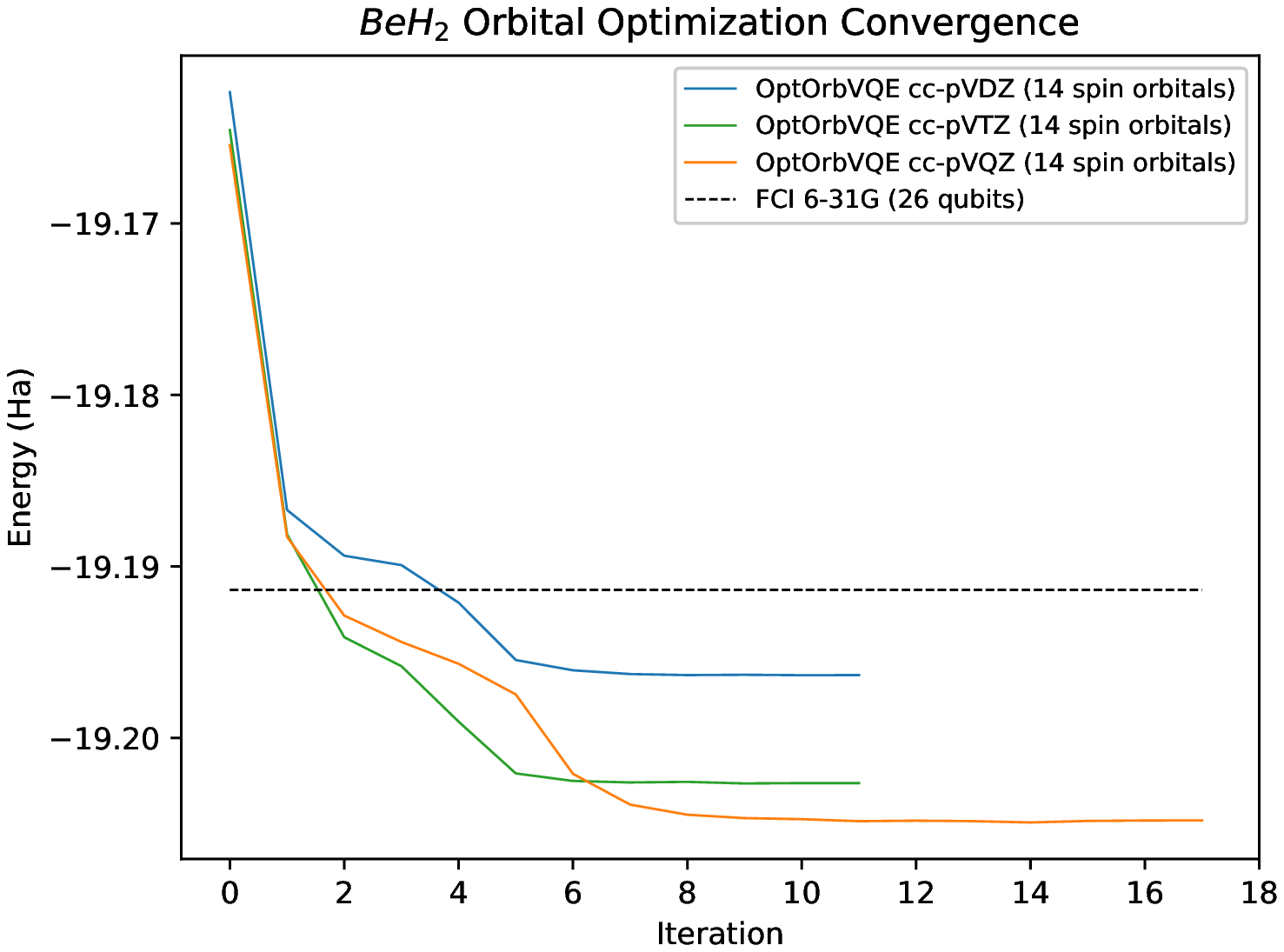}
    \caption{Convergence of OptOrbVQE as a function of the outer loop
    iteration number for \ch{BeH2} at the near-equilibrium \ch{Be}-\ch{H} bond
    distance of 1.3264 \AA.}
    \label{fig:BeH2_equilibrium}
\end{figure}
Fig.~\ref{fig:BeH2_equilibrium} illustrates the convergence of OptOrbVQE at
the equilibrium configuration. We find that starting from either the cc-pVTZ
or cc-pVQZ basis set results in OptOrbVQE surpassing the FCI energy in the
6-31G basis at the $n = 2$ iteration. Further iterations result in improved
energy. Starting from the cc-pVDZ also surpasses the FCI (6-31G basis),
but requires more iterations to do so.

\subsubsection{\ch{H2O}}\label{sec: H2O}

\begin{figure}[htb]
    \centering
    \includegraphics[width=\linewidth]{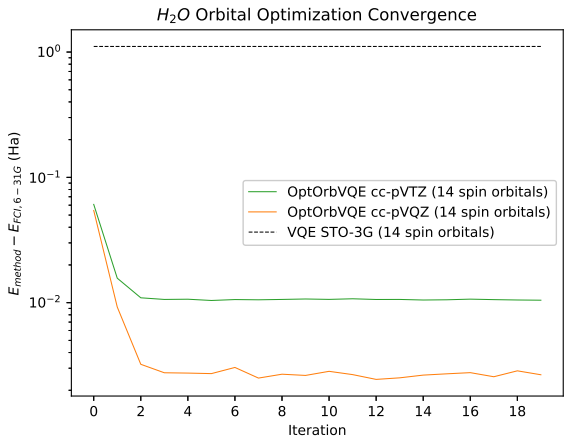}
    \caption{Convergence of OptOrbVQE as a function of the outer loop iteration
    number for \ch{H2O} at the near-equilibrium \ch{O}-\ch{H} bond distance
    of 0.9578 \AA~and bond angle 104.4776 degrees.} \label{fig:H2O_equilibrium}
\end{figure}

In this section we test OptOrbVQE on the \ch{H2O} molecule. The
ansatz used is 1-UCCSD. The number of spin orbitals used by OptOrbVQE is
set to 14, the number of spin orbitals for this molecule in the minimal
STO-3G basis. Fig.~\ref{fig:H2O_equilibrium} plots the difference of the
OptOrbVQE energy from the FCI energy in the 6-31G basis for \ch{H2O} at
the near-equilibrium configuration of O-H distance 0.9578 \AA~and H-O-H
bond angle of 104.4778 degrees. These results are different from the other
systems presented in that while the method still easily outperforms VQE
using the same number of spin orbitals, we do not observe OptOrbVQE using
a minimal number of spin orbitals to surpass the FCI energy in the larger
6-31G basis. OptOrbVQE can however be observed to approach the FCI (6-31G
basis) energy at the milli-hartree level, with the energy difference
converging to approximately $2.5\times10^{-3}$ Hartree when using cc-pVQZ
as the starting basis. One notable feature about this convergence curve is
that the rate of convergence is most rapid up until the $n = 3$ iteration,
then hits a plateau. The energy then fluctuates until the maximum number
of iterations is reached, indicating the possible presence of multiple
local minima which differ in energy at the milli-Hartree level. A similar
trend is observed when starting from the cc-pVTZ basis, although the
converged energy accuracy is worse and the fluctuations are less
pronounced in this case. It is also worth noting that the $0th$ iteration
of OptOrbVQE outperforms VQE in Fig.~\ref{fig:H2O_equilibrium}. Because
the initial partial unitary for OptOrbVQE is set to be the matrix which
selects the $N$ lowest energy spin orbitals, the $0th$ iteration
corresponds to starting with a large basis, but reducing the active space
to one the same size as the STO-3G basis. Thus, using orbital optimization
is often not necessary to outperform VQE in the STO-3G basis. The main
benefit of orbital optimization is further accuracy improvements at the
milli-Hartree level.

\subsection{Increasing Qubit Resources}

\begin{figure}[htb]
    \centering
    \includegraphics[width=\linewidth]{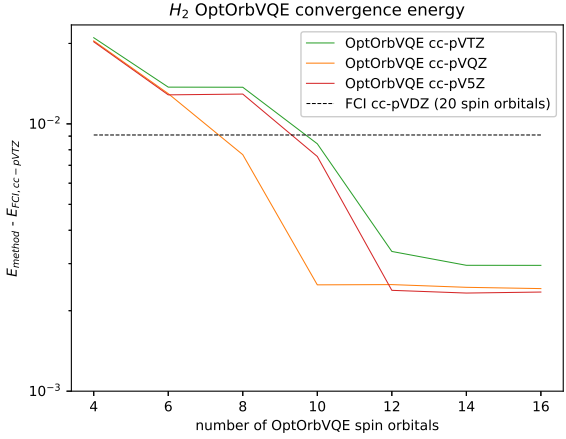}
    \caption{Converged energy of OptOrbVQE as a function of the number of
    spin orbitals for \ch{H2} at the near-equilibrium bond distance of
    0.735~\AA.}
    \label{fig:H2_increasing_qubits}
\end{figure}

One important feature of OptOrbVQE is that the number of spin orbitals used
is a tunable parameter that can be set to any positive integer up to the
number used by the starting basis set. The previous sections examined the
performance of OptOrbVQE for various systems when using a number of spin
orbitals equal to the minimal STO-3G basis. In this section we increase the
number of spin orbitals used by OptOrbVQE in order to examine the potential
for the method to further improve energy accuracies as the capabilities of
quantum computers improve with time. We test OptOrbVQE on \ch{H2} using even
integer numbers of spin orbitals from 4 to 16.  Qiskit's \emph{AerSimulator}
and \emph{AerPauliExpectation} are used to obtain ideal noiseless results
as in \S\ref{sec: Minimal Qubit Usage}. The optimizer used is L-BFGS-B and
the ansatz used is 1-UCCSD.  Fig.~\ref{fig:H2_increasing_qubits} plots the
difference of the OptOrbVQE energy at the near-equilibrium bond distance of
0.735~\AA~using OptOrbVQE and the FCI energy in the cc-pVTZ basis (56 spin
orbitals). The FCI energy in the cc-pVDZ basis (20 spin orbitals) is also
included for reference. The most significant (but expected) feature of this
plot is that the energy accuracy obtained by OptOrbVQE can be improved by
increasing the number of spin orbitals that it uses. This comes with the
caveat that using more qubits does not always result in a lower converged
energy. Several plateaus can be seen over the interval considered. For
example, increasing the number of spin orbitals from 6 to 8 does not result in
significantly improved energy when starting from either the cc-pVTZ or cc-pV5Z
basis sets. Increasing the number of qubits from 10 to 16 also does not appear
to result in improved energies when starting from the cc-pVQZ basis. Another
notable feature of this plot is that for a given number of qubits, starting
from a larger basis set does not always result in lower energy. This can be
seen from OptOrbVQE starting from the cc-pVQZ basis achieves a lower energy
than starting from the cc-pV5Z basis for 8 and 10 qubits. Finally, we note that in Fig.~\ref{fig:H2_increasing_qubits}, the green curve compares the logarithmic difference between the energy obtained by OptOrbVQE starting from the cc-pVTZ basis and the FCI energy in the full 56 spin-orbital cc-pVTZ basis. The highest degree of accuracy obtained at 16 spin-orbitals is approximately 3 milliHartree. There are several factors contributing to this discrepancy: \textbf{1.} OptOrbVQE consists of two optimization subproblems, neither of which is guaranteed to converge to the global minimum. Each may converge to a spurious local minima or within a neighborhood of the global minimum. \textbf{2.} The VQE subproblem utilizes a wavefunction ansatz. This comes with an associated ansatz representation error that is not present in the classical FCI algorithm. \textbf{3.} It is well-known that large basis set expansions improve the ability of computational methods to capture energy contributions that arise from electron correlation effects, in particular dynamic correlation. Although orbital optimization helps the method capture some of this energy contribution, its smaller basis size precludes it from capturing all of it.

The first two points listed here may also help to explain some un-intuitive behavior exhibited by some of the tests in this paper. For example, in Fig.~\ref{fig:LiH_equilibrium}, the largest starting basis used by OptOrbVQE, cc-pVQZ, is the one which achieved the least accurate energy among the three starting basis sets considered. This is counter to what one would intuitively expect, where the more flexible variational space should give it the potential to achieve the highest quality accuracy. Furthermore, several plateaus are observed for all three starting basis sets. Similarly, in Fig.~\ref{fig:H2_increasing_qubits} there are several instances where increasing the size of the variational space through an increase in the number of qubits does not strictly result in an increase in accuracy, but rather appears to occasionally result in a plateau. There are a few possible explanations for this behavior. We note that in order for the benefits of an increased variational space to be apparent in the final accuracy obtained, it is necessary for both the orbital optimization and VQE subproblems to converge sufficiently close to their global minima and for the VQE ansatz to have sufficient representation accuracy in the rotated basis sets determined by the orbital optimization subroutine at each iteration. If any of these conditions are not met, the final energy accuracy may not reach its full potential. We defer a more in-depth study on how to improve the convergence of OptOrbVQE to future work. For example, one could investigate incorporating adaptive ansatz strategies~\cite{Grimsley2019,PRXQuantum.2.020310} into the VQE subproblem. The intuition behind this approach is that an adaptive ansatz may be better suited for representing the ground state of a system than a fixed ansatz when the basis set representation itself is iteratively changing. A second possibility would be to add a random perturbation to the initial parameters of each VQE iteration. In these tests, a random perturbation is added to the initial partial unitary to help the orbital optimization escape from shallow local minima, but the VQE subproblem may also benefit from a similar initialization.

\subsection{Robustness to Noise}\label{sec: Robustness to Noise}

We now investigate the robustness of the OptOrbVQE algorithm to noise,
which we carry out in two stages using the binding curve of the \ch{H2}
molecule as a test system. In \S\ref{sec: Statistical Sampling Noise} we
incorporate statistical sampling as the only source of the noise. On quantum
hardware, this type of noise arises from the repeated circuit preparation
and observable measurement process. For example, to measure the quantity
$\bra{\psi_{\text{ref}}}\hat{U}^{\dagger}(\pmb{\theta})\hat{H}\hat{U}(\pmb{\theta})\ket{\psi_{\text{ref}}}$
we would prepare the circuit $\hat{U}(\pmb{\theta})$ $n$ times, measuring
each of the Pauli terms $\hat{P}_i$ in Eq.~\eqref{eq: qubit Hamiltonian}
$n$ times and classically compute the weighted sum of their expectation
values. Because this form of noise is independent of the ansatz circuit
depth, starting with this form of noise allows us to compare OptOrbVQE using
a smaller basis to VQE using a larger basis while keeping the effects that
would arise from the difference in circuit depth between these two problem
instances separate. In \S\ref{Depolarizing Noise} we add a local depolarizing
noise model to the statistical noise.

\subsubsection{Statistical Sampling Noise}\label{sec: Statistical Sampling Noise}

\begin{figure}[htb]
    \centering
    \includegraphics[width=\linewidth]{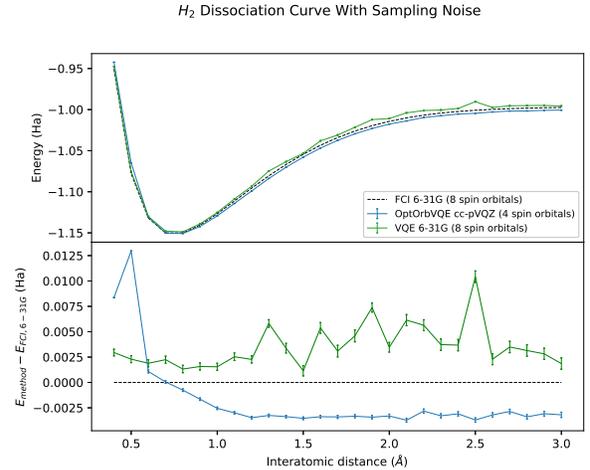}
    \caption{\textit{Top}: binding curve of \ch{H2} using $10^6$ circuit
    samples. \textit{Bottom}: difference in energy from the FCI (6-31G basis)
    energy.} \label{fig:H2_sampling_noise_binding_curve}
\end{figure}

For the noisy simulations, we choose COBYLA as the classical optimizer. Its
lack of a need to calculate gradient information makes it more resilient
to noise than L-BFGS-B. The ansatz used is 1-UCCSD. The mapping used
is Jordan-Wigner. $10^{6}$ circuit samples are used for observable
measurements. OptOrbVQE is set to use cc-pVQZ as the starting basis
and uses 4 spin orbitals in the transformed basis. We compare it to VQE
in the 6-31G basis (8 spin orbitals), using the FCI (6-31G basis) as a
baseline. Fig.~\ref{fig:H2_sampling_noise_binding_curve} illustrates the
results obtained for these tests. The outer loop stopping tolerance is set to
$10^{-3}$. The error bars are calculated internally by Qiskit, which records
the statistical variance $\sigma$ associated with expectation values from $n$
circuit samples and returns the error as $\sqrt{\frac{\sigma}{n}}$.

We can see that in the presence of statistical sampling noise, OptOrbVQE
retains its ability to achieve a lower ground state energy than VQE while only
using half the number of qubits for interatomic distances 0.6~\AA~and greater.

\subsubsection{Depolarizing Noise}\label{Depolarizing Noise}

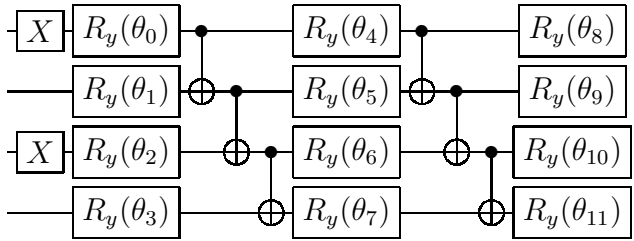
\begin{figure}[H]
    \centering
    \centerline{\Qcircuit @C=0.3em @R=.3em {
                        & \gate{X} &\gate{R_{y}(\theta_{0})} & \ctrl{1} & \qw & \qw & \gate{R_{y}(\theta_{4})} & \ctrl{1} & \qw & \qw & \gate{R_{y}(\theta_{8})} \\
                        & \qw &\gate{R_{y}(\theta_{1})} & \targ & \ctrl{1} & \qw & \gate{R_{y}(\theta_{5})} & \targ & \ctrl{1} & \qw & \gate{R_{y}(\theta_{9})} \\
                        & \gate{X} &\gate{R_{y}(\theta_{2})} & \qw & \targ & \ctrl{1} & \gate{R_{y}(\theta_{6})} & \qw & \targ & \ctrl{1} & \gate{R_{y}(\theta_{10})} \\
                        & \qw &\gate{R_{y}(\theta_{3})} & \qw & \qw & \targ & \gate{R_{y}(\theta_{7})} & \qw & \qw & \targ & \gate{R_{y}(\theta_{11})}
}}
    \caption{Ansatz used for \ch{H2} simulations with depolarizing noise.}
    \label{fig:NoisyH2Ansatz}
\end{figure}

In order to model the effects of gate noise, we add a local depolarizing
channel to each one-qubit gate and a tensor product of two local depolarizing
channels to each two-qubit gate. This has the effect that every time a
one-qubit gate is applied, one of the three Pauli operators (with equal
likelihood) is also applied with probability $p_{error}$. For two-qubit gates,
this probabilistic error event occurs independently for each qubit involved. In
this work we set $p_{error} = 10^{-3}$. No error mitigation techniques are
used. Aside from adding gate noise, the methodology remains the same as in
\S\ref{sec: Statistical Sampling Noise}, except that the ansatz is changed from
1-UCCSD to a hardware-efficient ansatz shown in Fig.~\ref{fig:NoisyH2Ansatz}.

\begin{figure}[htb]
    \centering
    \includegraphics[width=\linewidth]{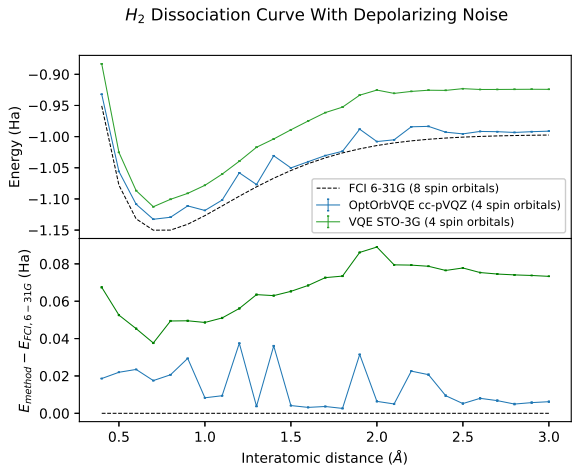}
    \caption{\textit{Top}: binding curve of \ch{H2} using $10^6$
    circuit samples with $p_{error} = 10^{-3}$. \textit{Bottom}:
    difference in energy from the FCI (6-31G basis) energy.}
    \label{fig:H2_depolarizing_noise_binding_curve}
\end{figure}

In Qiskit, this corresponds to the \emph{Real Amplitudes} circuit with the
number of repetitions set to 2. The first layer of this circuit prepares
the qubits in the Hartree Fock state. The parameters are initialized to
zero. We compare OptOrbVQE to VQE (STO-3G basis), using the FCI (6-31G
basis) energy as a baseline. The results of these tests are shown in
Fig.~\ref{fig:H2_depolarizing_noise_binding_curve}.

We find that OptOrbVQE consistently achieves lower energy than VQE when using
the same number of qubits. Unlike in \S\ref{sec: Statistical Sampling Noise}
when only statistical sampling noise was used, OptOrbVQE no longer achieves
energy lower than FCI in the 6-31G basis. It does, however, approach this
reference energy at the milli-Hartree level for several interatomic distances.

\section{Discussion and Conclusions}\label{sec: Discussion and Conclusions}

One of the main challenges that exists today in quantum computing is
demonstrating quantum advantage on a problem with practical utility. One
such problem is calculating the ground state of electronic chemical systems
to high accuracy when compared to laboratory results. In this work we have
demonstrated that OptOrbVQE offers a clear path towards this goal in two
ways: \textbf{1.} When using a number of qubits equal to that in a minimal
basis, OptOrbVQE consistently achieves higher accuracy than VQE using a
minimal basis set. In many cases it can even outperform VQE methods
of larger basis sets using a fraction of the number of qubits. As an aside, we also note that because these numerical demonstrations used an initial partial unitary which selects the subset of spin-orbitals with the lowest Hartree-Fock energy, the 0th OptOrbVQE energy is equivalent to that which would be obtained by VQE using an active space selected in this manner. Thus, we observe that OptOrbVQE achieves more accurate results than VQE when the starting underlying full orbital space basis is the same as well.  \textbf{2.}
The number of qubits used by OptOrbVQE is a tunable parameter.  Increasing the
number of qubits typically has the effect of improving the energy accuracy,
which provides a convenient method for systematically demonstrating improved
results as the capabilities of quantum computers progress.

This improved performance comes at the cost of running the orbital optimization
and VQE subproblems multiple times. While we find that our classical
simulations can in some instances utilize 10 or more iterations before the
stopping condition is reached, the bulk of the convergence typically occurs
during the first 2-5 iterations. These first few iterations are typically
sufficient for the method to surpass VQE and FCI methods of larger basis
sets. A user of this algorithm could simply choose to limit the number of
iterations to 2-5 and still see most of the benefit of this method over
using VQE with a basis set of the same size or larger.

One final point to note is that although we have used VQE to demonstrate
this method, many other quantum eigensolvers could be used in its place to
achieve different goals or to improve the performance. The main criterion
is that the eigensolver returns an improved estimate for the eigenstate(s)
over its input state(s). For example, Quantum Phase Estimation
(QPE)~\cite{doi:10.1126/science.1113479} would not be a suitable choice of
eigensolver because it returns an estimate of the eigenvalue but does not
return an improved estimate of the eigenstate itself. However, an
algorithm such as $\alpha$-VQE~\cite{PhysRevLett.122.140504} that uses QPE
as a subroutine could be a suitable eigensolver because it iteratively
improves the estimation of the ground state. Other suitable ground state
eigensolvers that could be explored in this orbital optimization framework
include Quantum Imaginary Time Evolution (QITE)~\cite{Motta2020},
variational QITE~\cite{McArdle2019}, Quantum Monte
Carlo~\cite{Huggins2022}, ADAPT-VQE~\cite{Grimsley2019}, and
qubit-ADAPT-VQE~\cite{PRXQuantum.2.020310}. Excited state eigensolvers
could be explored as well. The three most obvious candidates would be
Quantum Subspace Expansion (QSE)~\cite{PhysRevA.95.042308}, quantum
Equation of Motion (qEoM)~\cite{PhysRevResearch.2.043140}, and EOM-VQE~\cite{EOM-VQE}. These methods
operate by first performing the ground state search using an algorithm
such as VQE, then performing a classical post-processing diagonalization
step to find low-lying excited states of the Hamiltonian. Thus, OptOrbVQE
could be used as a ground state solver for these methods. Two other
excited states eigensolver for which it would be straightforward to
incorporate this orbital optimization procedure would be multistate
contracted VQE (MC-VQE)~\cite{PhysRevLett.122.230401} and Subspace Search
VQE (SSVQE)~\cite{PhysRevResearch.1.033062}. These two methods both apply
an ansatz circuit to a set of mutually orthogonal input states and
minimize an objective function consisting of a weighted sum of expectation
values of the Hamiltonian with respect to each of the resulting
parameterized states. OptOrbVQE could easily be generalized to
``OptOrbMC-VQE'' or ``OptOrbSSVQE'' by modifying Eq.~\eqref{eq: total
minimization problem} to be a weighted sum of the transformed Hamiltonian
with respect to mutually orthogonal parameterized states in the same
manner as these methods. Orbital optimization could also be applied to the
quantum Orbital Minimization Method
(qOMM)~\cite{doi:10.1021/acs.jctc.2c00218} by modifying Eq.~\eqref{eq:
total minimization problem} in an analogous way. These methods all find
low-lying excited states simultaneously through the minimization of a
single objective function. Variational Quantum Deflation
(VQD)~\cite{Higgott2019variationalquantum} is different from these other
methods in that it finds the low-lying excited states sequentially through
a series of minimization procedures. Thus, the application of orbital
optimization to VQD would be more involved than simply modifying
Eq.~\eqref{eq: total minimization problem}, but could still be
investigated. We leave the investigation of the application of the orbital
optimization procedure to these eigensolvers to future work.

\begin{acknowledgement}
    The work is supported in part by the US National Science Foundation
    under award CHE-2037263 and by the US Department of Energy via grant
    DE-SC0019449. YL is supported in part by the National Natural
    Science Foundation of China 12271109.
\end{acknowledgement}

\bibliography{reference}
\end{document}